\documentclass[lnicst]{svmultln}
\usepackage{makeidx}  
\usepackage{psfig, url,amssymb,amsmath}
\begin{document}
\mainmatter              
\title{Why the Internet is so `small'?}
\titlerunning{Internet Topology}  
%
\author{Shi Zhou}
\authorrunning{Shi Zhou}   

\institute{Department of Computer Science, University College London\\Malet Place, London, WC1E 6BT, United Kingdom\\
\email{s.zhou@cs.ucl.ac.uk}}

\maketitle

\begin{abstract}

During the last three decades the Internet has experienced fascinating evolution, both exponential growth in
traffic and rapid expansion in topology. The size of the Internet becomes enormous, yet the network is very
`small' in the sense that it is extremely efficient to route data packets across the global Internet. This
paper provides a brief review on three fundamental properties of the Internet topology at the autonomous
systems (AS) level. Firstly the Internet has a power-law degree distribution, which means the majority of
nodes on the Internet AS graph have small numbers of links, whereas a few nodes have very large numbers of
links. Secondly the Internet exhibits a property called disassortative mixing, which means poorly-connected
nodes tend to link with well-connected nodes, and vice versa. Thirdly the best-connected nodes, or the rich
nodes, are tightly interconnected with each other forming a rich-club. We explain that it is these structural
properties that make the global Internet so `small'.
\keywords{Internet, network, topology, autonomous systems, BGP, shortest path, power-law, scale-free,
assortative mixing, rich-club}
\end{abstract}

\section{Introduction}

The Internet is a network of autonomous systems (AS) which are collections of IP networks and routers under
the control of one entity, typically an Internet service provider. The Border Gateway Protocol (BGP) is a
critical component of the Internet's infrastructure as it serves to connect these ASes
together~\cite{quoitin03}. Performance of the BGP depends strongly on properties of the topological
connectivity between these ASes. Surprisingly, it was only in 1999 that researchers reported that the
Internet AS graph is a `scale-free' network~\cite{Barabasi99}. This discovery effectively invalidated all previous
Internet models based on random graphs.

Since then researchers have reported many other topological properties of the Internet~\cite{pastor04, mahadevan05b}. In this paper we
review three of them, namely the power-law degree distribution~\cite{Faloutsos99}, the disassortative mixing~\cite{Pastor01, newman03} and the rich-club
phenomenon~\cite{zhou04a}. We explain that it is these structural properties that make the global Internet so `small' in the
sense that it is extremely efficient to route data packets across the global Internet.

So far much of the research on Internet topology occurs in the communities of physics, mathematics and
complexity science. Interdisciplinary communication with other communities is much needed~\cite{Krioukov07}. This paper is an
effort towards this direction. The network under study is the Internet, but the statistical methods and the
topological properties introduced here can be applied to any other network in nature and society.

\section{Internet Topology at AS Level}

\begin{figure}[tbh]
\centerline{\psfig{figure=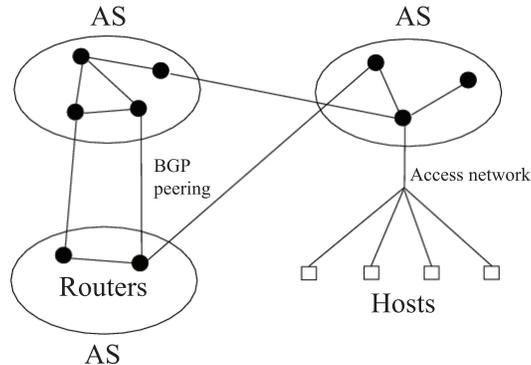,width=7cm}}
\caption{\label{fig:internet} Nodes of the Internet
topology. (1) Hosts, that are the computers of users; (2) routers, that arrange data traffic across the
Internet; and (3) autonomous systems (AS) that are subnetworks in the Internet.}
\end{figure}

The Internet is a capital example of complex networks. It is unlike any previous human invention in both
scale and effect, and is now a global resource important to all of the people in the world. The Internet
provides a low-level communication infrastructure upon which other communication mechanisms can be built,
e.g. the ubiquitous email and Web protocols. Internet service providers (ISP) offer billions of users access
to the Internet via telephone line, cable and wireless connections. Data packets generated by users are
forwarded across the Internet toward their destinations by routers through a process known as routing.

The Internet contains millions of routers, which are grouped into thousands of subnetworks, called autonomous
systems (AS) (see Fig.~\ref{fig:internet}). The global Internet structure is characterised at the AS-level
because the delivery of data traffic through the Internet depends on the complex interactions between ASes
that exchange routing information using the Border Gateway Protocol (BGP). On the Internet AS-level topology,
a node is an AS, which is usually controlled by an ISP; and a link represents a BGP peering relation between
two ASes, which a commercial agreement between two ISPs on forwarding data traffic.

Measurement data of the Internet AS-level topology became available in late
1990's. Since then there have been a number of projects to collect the topology data of the Internet~\cite{mahadevan05b}. These
measurements have greatly improved our understanding of the structure and evolution of the
Internet~\cite{pastor04}. The Internet macroscopic structure has become relatively stable in recent years.
This suggests that the Internet has entered a fairly mature stage in terms of network evolution. Today the
Internet AS graph is very large. It contains about twenty thousand AS nodes and fifty thousands of links. It
is important to realise that the Internet is actually very much sparsely connected. The actual number of
links present in the Internet is only about 0.04\% of all possible links that the network can have. This is
relevant because the fewer links the Internet has the cheaper and easier to invest, manage and maintain the
network.

A remarkable property of the Internet is that it is extremely efficient in routing data packets across the
global network. On the AS graph the average of shortest path between any two nodes is around 3 hops. The
length of routing path observed on real BGP data is around 4 hops. This is in spite of the fact that the
Internet contains tens of thousands of nodes and the network is very sparsely connected. Why is the Internet
so `small'? The key lies in the topological structure of the Internet.

\section{Node Connectivity}

In graph theory, degree $k$ is defined as the number of links or immediate neighbours a node has.  The average node degree is $\langle k
\rangle={1\over N}\sum_i k_i =2L/N$, where $k_i$ is the degree of node $i$, $N$ is the number of nodes and
$L$ is the number of links. The average node degree in the Internet is around 6.
The principal statistical property  to
describe and discriminate between different networks is to measure the degree distribution~$P(k)$, i.e.~the
fraction of nodes in the network with degree~$k$.

It had been assumed that the Internet topology resembles a random graph where nodes are connected with each
other randomly (with a uniform probability). A random network features a Poisson degree distribution where
most node degrees are close to the average degree and only a few nodes  have very large or very small degrees. It is known from graph theory that a random graph is small in the sense that one node is never too far away from other nodes. Is this the
reason that the Internet is so small?

\begin{figure}[tbh]
\centerline{\psfig{figure=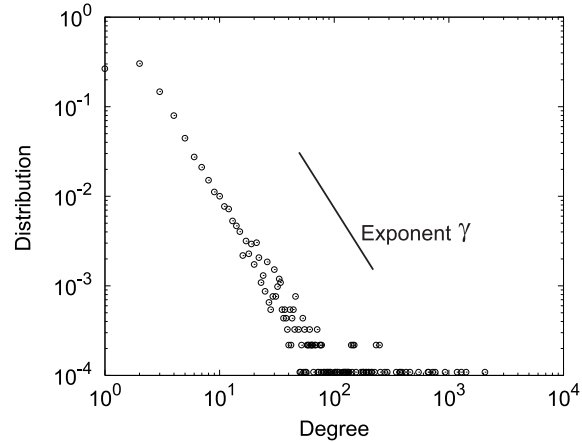,width=8cm}} \caption{\label{fig:evaluation:PK}Internet
power-law degree distribution on log-log scale.}
\end{figure}

The answer is no, because in 1999 it was discovered that the internet topology at the AS level (and at the
router level) exhibits a power law degree distribution $P(k)\sim C k^{\gamma}$ \cite{Faloutsos99}, where $C >
0$ is a constant and the exponent $\gamma\simeq -2.2\pm0.1$ (see Fig.\,\ref{fig:evaluation:PK}). This means a
few nodes have very large numbers of connections, whereas the vast majority of nodes have only a few links.
The power law property is an evidence that the Internet AS-level topology has evolved into a very heterogeneous
structure, which is completely different from the random graph model. A power-law network is called a
`scale-free' network~\cite{Barabasi99} because it is not the average degree, but the exponent of power-law
distribution that fundamentally characterises the network's connectivity and the value of the exponent is not related to the size (scale) of the network.

Recently the small-world theory~\cite{watts98} has attracted a lot of attention. A small-world network resembles a typical
social network which is highly clustered where one's friends are likely to be friends to each other. At the same time a small-world contains random connections which serve as shortcuts and make the network small.  Most
scale-free networks can be regarded as small-world networks in the sense that the average shortest paths
of scale-free networks are always small. But some scale-free networks are smaller than others. For example the average
shortest path of many social networks is around 6 (which is comparable to that of random graphs);
whereas the Internet's average shortest path is 3 which is actually substantially smaller than predicted by the random graph and
small-world theories.

In the last few years many studies have shown~\cite{ mahadevan05b, zhou07b} that the degree distribution alone does
not uniquely characterise a network topology. Networks having exactly identical degree distribution can
exhibit vastly different other topological properties. The fact that a network is scale-free only tells us limited information. In order to obtain a full picture of a network's structure,
we need to look beyond the question of `how many links does a node have?' and ask `To whom the node is
connected to?'

\section{Who Connects with Whom?}

\subsection{Disassortative Mixing}

Recently a number of studies have shown that the degree correlation plays a significant role in defining a network's structure~\cite{mahadevan05b,
newman03, vazquez03, maslov04}. The degree correlation defines the mixing pattern in a network. Social
networks are assortative mixing where nodes tend to attach to alike nodes, i.e.~high-degree nodes to
high-degree nodes and low-degree nodes to low-degree nodes. On contrast technological and biological
networks, including the Internet, are disassortative mixing where high-degree nodes tend to connect with
low-degree nodes, and visa versa. We can not use a network's degree distribution to predict its mixing pattern. Networks with the same degree distribution can have completely opposite mixing patterns~\cite{zhou07b}.

\begin{figure}[tbh]
\centerline{\psfig{figure=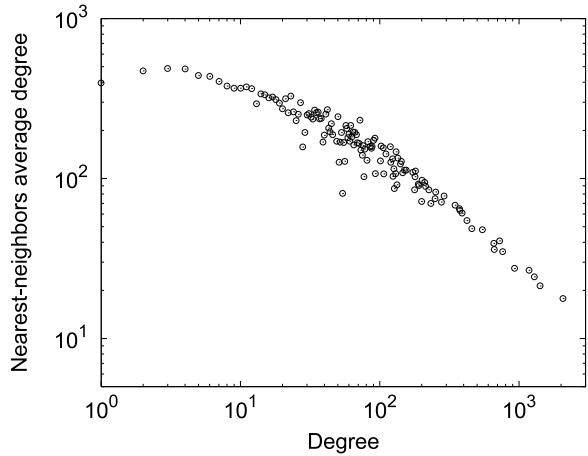,width=8cm}}
\caption{\label{fig:evaluation:Knn}Nearest-neighbours average degree of $k$-degree nodes. }
\end{figure}

A network's mixing pattern can be inferred by the correlation between a node's degree $k$ and its
nearest-neighbours average degree $k_{nn}$~\cite{Pastor01}. Fig.~\ref{fig:evaluation:Knn} shows the Internet
exhibits a negative correlation between the two quantities suggesting the network is disassortative mixing.
For example for a node with degree 1000, the average degree of its 1000 neighbours is around 20; whereas for
a node with degree 2, the average degree of its two neighbours is around 500.

The mixing pattern can also be inferred by computing the so-called assortative coefficient~\cite{newman03},
$-1<\alpha<1$, which is defined as
$$
\alpha = {L^{-1}\sum_{m} j_m k_m - [L^{-1}\sum_m{1\over2}(j_m+k_m)]^2 \over
L^{-1}\sum_m{1\over2}(j_m^2+k_m^2)-[L^{-1}\sum_m{1\over2}(j_m+k_m)]^2},
$$
where $L$ is the number of links a network has, and $j_m$, $k_m$ are the degrees of the nodes at the ends of
the $m$th link, with $m=1,...,L$. If $0<\alpha<1$, a network is assortative mixing; and if $-1<\alpha<0$, a
network is disassortative mixing, e.g.~$\alpha=-0.19\sim-0.24$ for the Internet.

The disassortative mixing property of the Internet is relevant because it means the peripheral, low-degree
nodes are almost always connect directly with some high-degree nodes. On the other hand the chance of finding a chain of
low-degree nodes is low. Then, how about the high-degree nodes themselves? Are they interconnected with each
other? Does the disassortative mixing property mean that the high-degree nodes only connect with low-degree nodes?

\subsection{Rich-Club Phenomenon}

\begin{figure}
\centerline{\psfig{figure=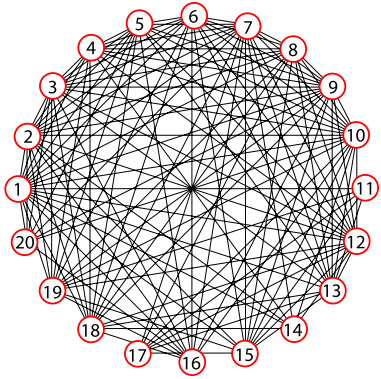,width=5cm}} \caption{\label{fig:richclub} Internet connections
among the top 20 richest nodes themselves. Other connections to the nodes are omitted for clarity.}
\end{figure}

We call nodes with large numbers of links the `rich' nodes. Fig.\,\ref{fig:richclub} shows that in the
Internet the richest nodes are tightly interconnected with each other forming the core of the network. This is
called the rich-club phenomenon~\cite{zhou04a}. This phenomenon does not conflict with the Internet's
disassortative mixing property. Each rich node shown in the figure has a large number of links. The majority
the links are connect to poorly connected nodes (disassortative mixing) whereas  only a few of them are
enough to provide the interconnectivity with other rich nodes, whose number is anyway small.

\begin{figure}
\centerline{\psfig{figure=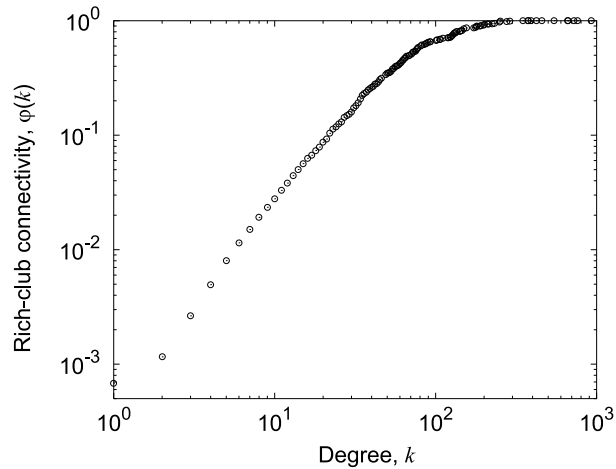,width=8cm}}
\caption{\label{fig:evaluation:rich-degree}Rich-club coefficient as a function of node degree.}
\end{figure}

The rich-club phenomenon can be quantitatively assessed by measuring the rich-club coefficient,
\begin{equation}
\phi(k)=\frac{2E_{\geqslant k}}{N_{\geqslant k}(N_{\geqslant k} - 1)},
\end{equation}
where $N_{\geqslant k}$ is the number of nodes with degrees greater or equal to $k$ and $E_{\geqslant k}$ is
the number of links among the $N_{\geqslant k}$ nodes themselves. The rich-club coefficient measures the
ratio of the actual number of links  to the maximum possible number of links among nodes with degrees no less
than $k$.  Fig.~\ref{fig:evaluation:rich-degree} shows  the rich-club efficient as a function of node degree
$k$. It shows that the rich nodes are becoming more tightly interconnected when the degree increases, in other words, when the club becomes more `exclusive'. A small number of nodes with
degrees larger than 300 have $\varphi=1$ which means they are fully interconnected.

\section{Discussion}

The internet is ultra small because it exhibits both disassortative mixing and rich-club phenomenon. These two
structural properties together contribute to the routing efficiency of the network. The rich club consists of
a small number of highly connected nodes. The club members are tightly interconnected between each other. If
two club members do not have a direct connection, they are very likely to share a neighbouring club member.
Thus the hop distance between the members is very small(either 1 or 2 hops). The rich-club functions as a
`super' traffic hub of the Internet by providing a large selection of shortcuts for routing. The
disassortative mixing property ensures that the majority of network nodes, which are peripheral low-degree
nodes, are always near the rich-club. Thus a typical shortest path between two peripheral nodes consists of
three hops, the first hop is from the source node to a member of the rich club, the second hop is between two
rich-club members, and the final hop is to the destination node.

As we know a fully connected graph is the most efficient topology
because the shortest path is 1 between any two nodes. Whereas for sparsely connected networks, the star-like
topology is very efficient because the average shortest path is less than 2. If we abstract the rich-club as a single node then the Internet topology will look like a star topology, where all
nodes connects with the centre of the star. A star topology has a big problem that the centre of the star is very easy to be congested and becomes the single-point-of-failure.
The Internet `solves' these problems by replacing the centre with a club of rich nodes. The rich-club
contains plenty of redundant links. The network's performance and integrity will not be affected if some of
the links between the club members are removed. In addition, incoming traffic to the `centre' is handled in a
distributed manner by a handful of rich-club members such that the chance of getting congested is reduced. In
case a few club member are congested (or removed), the peripheral nodes served by these rich nodes may suffer,
but the rest of the network can still work as normal. Of course if the rich-club core is broken by removing
too many rich nodes or inter-rich links, the network will fall apart.

\section{Summary}

In a way the Internet has an ideal structure. It is very large in size and very sparse in
connectivity. Yet the Internet has a sophisticated structure which makes the network very `small' in terms of
routing. At the same time the Internet is quite robust because the function of the network's core is shared
by a group of rich nodes which are tightly interconnected with each other and therefore have plenty of redundant
links. It is remarkable that such a structure is the consequence of the Internet evolution process which is
driven by distributed, local decisions without any central plan. In fact until  ten years ago there was not even any map of the global Internet .

In the last decade there is a increasing recognition that effective engineering of the Internet is predicated on a detailed
understanding of issues such as the large-scale structure of its underlying physical topology, the manner in
which it evolves over time, and the way in which its constituent components contribute to its overall
function~\cite{floyd03}.

In recent years there have been a lot of research activities on characterising and modelling the Internet
topology. Various metrics and  models are proposed~\cite{tangmunarunkit02, Zhou04d}. Researchers are interested in understanding not only the
Internet's structure but also the dynamics and the evolution~\cite{dorogovtsev03a}. One important goal is to provide better models for more realistic simulations which differentiates the different
natures of links between AS nodes and incorporates suitable traffic model as well as other relevant factors
such as capacity and demo-geographic distributions of network elements.

\subsubsection*{Acknowledgments.} S.~Zhou is supported by The
Royal Academy of Engineering/EPSRC Research Fellowship under grant no.~10216/70.


\end{document}